\documentclass[aps,floatfix,preprint,preprintnumbers,nofootinbib,superscriptaddress,natbib]{revtex4}
\usepackage{graphicx,float}% Include figure files
\usepackage{epsfig}		
\usepackage{dcolumn}% Align table columns on decimal point
\usepackage{bm}% bold math
\usepackage{subfigure}

\usepackage[all]{xy}
\usepackage{amsmath,upgreek}
\usepackage{amssymb}

\usepackage{pdfpages}
\usepackage{color}
\usepackage{graphicx,epstopdf}
\usepackage[colorlinks=true,
 linkcolor=red,
 urlcolor=purple,
 citecolor=blue,hyperindex]{hyperref}
\def\0{\mbox{\tiny $0$}}
\def\1{\mbox{\tiny $1$}}
\def\2{\mbox{\tiny $2$}}
\def\3{\mbox{\tiny $3$}}
\def\4{\mbox{\tiny $4$}}
\def\5{\mbox{\tiny $5$}}
\def\6{\mbox{\tiny $6$}}
\def\7{\mbox{\tiny $7$}}
\def\8{\mbox{\tiny $8$}}
\def\9{\mbox{\tiny $9$}}

\def\f14{\mbox{\tiny $\frac{1}{4}$}}

\begin{document}

\title{Phase-space quantum distorted stability pattern for Aubry-Andr\'e-Harper dynamics}

\renewcommand{\baselinestretch}{1.2}
\author{A. E. Bernardini}
\email{alexeb@ufscar.br}
%\altaffiliation[On leave of absence from]
\affiliation{~Departamento de F\'{\i}sica, Universidade Federal de S\~ao Carlos, PO Box 676, 13565-905, S\~ao Carlos, SP, Brasil.}
\author{O. Bertolami}
\email{orfeu.bertolami@fc.up.pt}
\altaffiliation[Also at~]{Centro de F\'isica do Porto, Rua do Campo Alegre 687, 4169-007, Porto, Portugal.} 
\affiliation{Departamento de F\'isica e Astronomia, Faculdade de Ci\^{e}ncias da Universidade do Porto, Rua do Campo Alegre 687, 4169-007, Porto, Portugal.}
\date{\today}

\begin{abstract}
Instability features associated to topological quantum domains which emerge from the Weyl-Wigner (WW) quantum phase-space description of Gaussian ensembles driven by Aubry-Andr\'e-Harper (AAH) Hamiltonians are investigated.
Hyperbolic equilibrium and stability patterns are then identified and classified according to the associated (nonlinear) AAH Hamiltonian parameters.
Besides providing the tools for quantifying the information content of AAH systems, the Wigner flow patterns here discussed suggest a systematic procedure for identifying the role of quantum fluctuations over equilibrium and stability, in a framework which can be straightforwardly extended to describe the evolution of similar/modified AAH systems.
\end{abstract}

%\pacs{03.65.-w, 03.65.Sq, 81.07.St, 03.65.Ta}

\date{\today}
\maketitle

\section{Introduction}

The threshold between periodic structures and disordered systems, including a full description of quasiperiodicity, is of broad interest in solid-state physics and has sparked discussions across multiple disciplines. These include multifractal states, localization transitions, and mobility edges \cite{Sokoloff,Kohmoto,Abe,Geisel,Guarneri,Piechon,Ketzmerick,Moura,Jeon,Boers,Schreiber,Yao,Yao2}, as well as stochastic processes with applications in biology \cite{Allen,Novo21C,Novo21D,Novo222,PP00}.
Within a broader framework, (non-)periodicity, (dis)order, and (in)stability are analyzed in the context of transport phenomena in classical and quantum systems described by Hamiltonian dynamics. These systems range from competition-induced chaos and molecular microscopy \cite{PP00,PP01,PP02,PP03,PP04} to diverse condensed-matter models \cite{Kramer,Lahini,Kraus,Scherg,Ni,Wang}. Notably, they encompass light and atom localization \cite{Roati,Fallani} under a deterministic quasiperiodic potential governed by the Aubry-Andr\'e-Harper (AAH) model, a one-dimensional lattice model for quasicrystals exhibiting localized and delocalized phases \cite{Aubry,Thouless,Xianlong22}.

The foundational Harper model \cite{Harper,HarperB} initially provided a tight-binding approximation for symmetric cubic crystals, modeling the effect of a uniform magnetic field on conduction-band metals. It was later extended via the Harper-Hofstadter generalization \cite{Harper,HarperB,NatHarper,RMPRMP} to describe chiral many-body states in lattice systems with a gauge field \cite{NatHarper}. Recent advances using laser-assisted tunneling techniques have enabled single-site and single-particle control, allowing for tunable chiral systems \cite{PRL-Harper,PRA-Harper19}, where particle numbers are adjusted atom-by-atom to manipulate lattice size, with the band structure forming Hofstadter's butterfly \cite{Harper,Harper02}.
Harper's tight-binding description is closely related to the Aubry-Andr\'e framework for aperiodic order in a one-dimensional lattice, exhibiting a delocalization-localization transition at a finite quasiperiodic potential \cite{Longhi}. Beyond applications in electronic localization \cite{Sarkar20} and magnetoresistance engineering \cite{Patra19}, the AAH model supports non-Hermitian extensions \cite{Longhi,Xianlong22} and plays a role in topological classifications \cite{PRBPRB2}. In particular, its Hamiltonian formulation has motivated research into order-disorder thresholds \cite{PREPRE}, potentially linked to quantum topological phenomena \cite{PRBPRB}.

Semiclassically, the AAH model can be mapped onto the effective Hamiltonian:
\begin{equation}\label{HamHarper01}
H^{W}_{_{AAH}}(q,\, p)= \pm \left[2\pi\beta\cos(p/p_0) + 2\pi\beta\cos(q/q_0)\right]+\omega_p p+\omega_q q,
\end{equation}
where an anisotropic $q-p$ phase-space has the corresponding analogous quantum mechanical operators, $\hat{q}$ and $\hat{p}$, obeying the commutation relations, $[\hat{q},\hat{p}] = i\,2\pi\beta\,p_0\,q_0$, with $\beta$ identified as the Peierls phase \cite{Harper,Harper02,PRA-Harper19}. This effectively works as a modulation of $\hbar$, with the periodicity of phase-space trajectories destroyed by momentum and position contributions driven by $\omega_p$ and $\omega_q$, leading to either bound-state or continuous spectrum regimes.

Given the above setup, our work examines equilibrium and stability conditions under quantum distortions in AAH systems. These conditions connect to stationarity and non-Liouvillianity within the phase-space Weyl-Wigner (WW) framework \cite{Wigner,Hillery,Case}. Since WW formalism provides a fluid-like description of phase-space information flow \cite{Steuernagel3,NossoPaper,Meu2018}, the classical and quantum information encoded in Eq.~\eqref{HamHarper01} can be mapped onto quantum phase-space ensemble dynamics. Wigner currents thus quantify the impact of quantum fluctuations on probability distributions and information flows \cite{NossoPaper,Steuernagel3,Meu2018}.

Building on previous studies of nonlinear ($q$, $p$) Hamiltonian systems \cite{Novo21A,Novo21B,Novo21C}, this approach establishes a platform for exploring microscopic-quantum and macroscopic-classical interplay. Non-equilibrium and instability features of topological quantum domains in phase space are identified through an autonomous system of ordinary differential equations governing the Wigner currents \cite{Novo21C,Novo21D,Novo222} such that stability properties, analyzed via hyperbolic equilibrium parameters, reveal the system's response to quantum distortions. Hence, the main proposal of this work is to compute equilibrium and stability quantifiers for quantum Gaussian ensembles driven by the effective AAH Hamiltonian, Eq.~\eqref{HamHarper01}, engendering a probe for quasiperiodicity patterns.
 
The manuscript is organized as follows: Sec. II reviews stationarity and Liouvillianity properties in the extended Wigner framework for $H^{W}(q,\,p) = K(p) + V(q)$ and their application to Gaussian ensembles. Sec. III examines the AAH system within the phase-space Wigner flow framework, ensuring full inclusion of quantum corrections. Sec. IV derives hyperbolic equilibrium and stability parameters, correlating them with (non-)stationarity and (non-)Liouvillianity via Wigner currents and ensemble parameters. Analytical expressions for Wigner currents are explicitly obtained. To conclude, Sec. V presents the outlook for future investigations.

\section{Extended Wigner framework}

The Weyl-Wigner (WW) \cite{Wigner,Hillery,Ballentine,Case} phase-space representation of quantum mechanics (QM) encompasses the dynamics of quantum systems in order to offer not only an enlarged view, but also an equivalent description of QM in terms of {\em quasi}-probability distribution functions of position and momentum coordinates.
Insights about the boundaries between quantum and classical physics as well as additional access to quantum information issues \cite{Neumann,Zurek01,Zurek02,Steuernagel3} are provided by the WW formalism without distorting the grounds of QM.
A connection between operator methods and path integral techniques \cite{Abr65,Sch81,Par88} is supported by the Weyl transform over a quantum operator, $\hat{O}$, described by
\begin{equation}
O^W(q,\, p)\label{Wigner111}
= 2\hspace{-.15cm} \int^{+\infty}_{-\infty} \hspace{-.35cm}ds\,\exp{\left[2\,i \,p\, s/\hbar\right]}\,\langle q - s | \hat{O} | q + s \rangle=2\hspace{-.15cm} \int^{+\infty}_{-\infty} \hspace{-.35cm} dr \,\exp{\left[-2\, i \,q\, r/\hbar\right]}\,\langle p - r | \hat{O} | p + r\rangle.
\end{equation} 
For $\hat{O}$ identified as a density matrix operator, $\hat{\rho} = |\psi \rangle \langle \psi |$, the Wigner {\em quasi}-probability distribution function is defined by
\begin{equation}
 h^{-1} \hat{\rho} \to W(q,\, p) = (\pi\hbar)^{-1} 
\int^{+\infty}_{-\infty} \hspace{-.35cm}ds\,\exp{\left[2\, i \, p \,s/\hbar\right]}\,
\psi(q - s)\,\psi^{\ast}(q + s),\label{Wigner222}
\end{equation}
which corresponds to the Fourier transform equivalent of the off-diagonal elements of $\hat{\rho}$, where $h = 2\pi \hbar$ is the Planck constant.
Such a definition provides a straightforward probability distribution interpretation constrained by the normalization condition over $\hat{\rho}$, $Tr_{\{q,p\}}[\hat{\rho}]=1$\footnote{In particular, this is consistent with the marginal distributions which, upon integration over canonical coordinates, return position and momentum distributions, i.e.
\begin{equation}
\vert \psi(q)\vert^2 = \int^{+\infty}_{-\infty} \hspace{-.35cm}dp\,W(q,\, p)
\qquad
\leftrightarrow
\qquad
\vert \varphi(p)\vert^2 = \int^{+\infty}_{-\infty} \hspace{-.35cm}dq\,W(q,\, p),
\end{equation}
constrained by the Fourier transform of the respective wave functions,
\begin{equation}
 \varphi(p)=
(2\pi\hbar)^{-1/2}\int^{+\infty}_{-\infty} \hspace{-.35cm} dq\,\exp{\left[i \, p \,q/\hbar\right]}\,
\psi(q).
\end{equation}}.

Of course, the vast scenario o WW phase-space QM is not constrained by the short set of the above properties as it also depicts the probabilistic interpretation from Schr\"odiger picture of QM, providing the tools for connecting averaged values with quantum observables \cite{Wigner,Case} as well as admitting extensions from pure states to statistical mixtures\footnote{
Considering the computation of average values by an overlap integral over the infinite volume described by the phase-space coordinates, $q$ and $p$, 
\begin{equation}
 \langle O \rangle = 
\int^{+\infty}_{-\infty} \hspace{-.35cm}dp\int^{+\infty}_{-\infty} \hspace{-.35cm} {dq}\,\,W(q,\, p)\,{O^W}(q,\, p), \label{eqfive}
\end{equation}
the replacement of ${O^W}(q,\, p)$ by $W(q,\, p)$ results into an analogous of the trace operation, $Tr_{\{q,p\}}[\hat{\rho}^2]$, which is identified by the quantum purity,
\begin{equation}
Tr_{\{q,p\}}[\hat{\rho}^2] = 2\pi\hbar\int^{+\infty}_{-\infty}\hspace{-.35cm}dp\int^{+\infty}_{-\infty} \hspace{-.35cm} {dq}\,\,\,W(q,\, p)^2.
\label{eqpureza}
\end{equation}}.

From a broader perspective \cite{Wigner,Hillery,Ballentine,Neumann,Zurek01,Zurek02}, the WW picture encompasses scenarios since from plasma and nuclear physics \cite{Sch,Lvovsky2009} up to quantum chaos \cite{Chaos} and quantum cosmology \cite{JCAP18}, including, for instance, the investigation of scattering and decoherence in the context of semiconductor transport process phenomenology \cite{Zachos2005,Jung09}. It has also been considered in the interpretation the wave-function collapse problem \cite{Zurek02,Bernardini13A,Leal2019}, in the paradigmatic extension of the standard QM \cite{Catarina,Bernardini13B,PhysicaA, Catarina001,Stein,Bernardini13C,Bernardini13E}, and as a background for parallel frameworks which include Husimi $Q$ \cite{Husimi,Ballentine} and Glauber-Sudarshan representations \cite{Glauber,Sudarshan,Carmichael,Callaway} of QM, and the highly specific optical tomographic probability framework \cite{Amosov,Radon,Mancini}.

More related to our work, the WW formalism supports a fluid equivalence of the phase-phase information flow \cite{Steuernagel3,NossoPaper,Meu2018}, from which the dynamical properties of the Wigner function, $W(q,\,p) \to W(q,\,p;\,t)$, can be straightforwardly related to the Hamiltonian dynamics.
A statistical inception of the density matrix operator, described in terms fo $W(q,\, p)$, maps a quantum phase-space ensemble dynamics so as to quantifiy quantum fluctuations which affect probability distributions and information flows \cite{NossoPaper,Steuernagel3}.
By construction, it results into a continuity equation written terms of a vector flux \cite{Steuernagel3,NossoPaper,Meu2018}, $\mathbf{J}(q,\,p;\,t)$, decomposed into position and momentum directions, $\mathbf{J} = J_q\,\hat{q} + J_p\,\hat{p}$, given by
\begin{equation}
{\partial_t W} + {\partial_q J_q}+{\partial_p J_p} =0,
\label{alexquaz51}
\end{equation}
with
\begin{equation}
J_q(q,\,p;\,t)= \frac{p}{m}\,W(q,\,p;\,t), \label{alexquaz500BB}
\end{equation}
and
\begin{equation}
J_p(q,\,p;\,t) = -\sum_{\eta=0}^{\infty} \left(\frac{i\,\hbar}{2}\right)^{2\eta}\frac{1}{(2\eta+1)!} \, \left[\partial_q^{2\eta+1}V(q)\right]\,\partial_p ^{2\eta}W(q,\,p;\,t),
\label{alexquaz500}
\end{equation}
where $\partial^s_a \equiv (\partial/\partial a)^s$ and the setup comes from a non-relativistic Hamiltonian operator, ${H}(\hat{Q},\,\hat{P})$, for which the Weyl transform yields 
\begin{equation}\label{Hamilt}
{H}(\hat{Q},\,\hat{P}) = \frac{\hat{P}^2}{2m} + V(\hat{Q}) \quad\to \quad H^{W}(q,\, p) = \frac{{p}^2}{2m} + V(q).
\end{equation}
The above set of equations reproduce a flow field dynamics\cite{Case,Ballentine,Steuernagel3,NossoPaper,Meu2018}, which can be reduced to the Liouville equation for the classical domain, with
quantum properties computed from higher order derivatives of the QM potential\footnote{The series expansion contributions from $\eta \geq 1$ introduce the quantum corrections which distort classical trajectories. The suppression of $\eta \geq 1$ contributions results into a classical Hamiltonian description of the phase-space probability distribution dynamics as given by
\begin{equation}
J^{\mathcal{C}}_q(q,\,p;\,t)= +({\partial_p H^{W}})\,W(q,\,p;\,t), \label{alexquaz500BB2}
\end{equation}
and
\begin{equation}
J^{\mathcal{C}}_p(q,\,p;\,t) = -({\partial_q H^{W}})\,W(q,\,p;\,t).
\label{alexquaz500CC2}
\end{equation}.} associated to nonlinear effects driven by powers of $h$. The fluctuations over the Liouvillian flow evince the threshold between classical and quantum regimes \cite{NossoPaper,JCAP18,Meu2018,Bernardini2020-02}.
On one hand, the classical phase-space velocity is given by $\mathbf{v}_{\xi(\mathcal{C})} = \dot{\mbox{\boldmath $\xi$}} = (\dot{q},\,\dot{p})\equiv ({\partial_p H^{W}},\,-{\partial_q H^{W}})$, with $\mbox{\boldmath $\nabla$}_{\xi}\cdot \mathbf{v}_{\xi(\mathcal{C})}= \partial_q \dot{q} + \partial_p\dot{p} = 0$, and where {\em dots} correspond to time derivatives, $d/dt$.
On the other hand, the quantum current can be parameterized by, $\mathbf{J} = \mathbf{w}\,W$, where the Wigner phase-space velocity, $\mathbf{w}$, is the quantum analog of $\mathbf{v}_{\xi(\mathcal{C})}$.
In this case, given that $\mbox{\boldmath $\nabla$}_{\xi}\cdot\mathbf{J} = W\,\mbox{\boldmath $\nabla$}_{\xi}\cdot\mathbf{w}+ \mathbf{w}\cdot \mbox{\boldmath $\nabla$}_{\xi}W$, one has \cite{Steuernagel3}
\begin{equation}
\mbox{\boldmath $\nabla$}_{\xi} \cdot \mathbf{w} = \frac{W\, \mbox{\boldmath $\nabla$}_{\xi}\cdot \mathbf{J} - \mathbf{J}\cdot\mbox{\boldmath $\nabla$}_{\xi}W}{W^2},
\label{zeqnz59}
\end{equation}
and Wigner function stationary and Liouvillian behaviors can be identified and quantified in terms of Eqs.~\eqref{alexquaz51} and \eqref{zeqnz59}, respectively, by $\mbox{\boldmath $\nabla$}_{\xi} \cdot \mathbf{J} = 0$ and $\mbox{\boldmath $\nabla$}_{\xi} \cdot \mathbf{w} = 0$ \cite{NossoPaper,JCAP18,Meu2018,Bernardini2020-02}.

Turning back to the analysis of the AAH dynamics, and considering results on the WW formalism \cite{Novo21A,Novo21B,Novo21C,Novo21D,Novo222}, the replacement of a Schr\"odinger-like dynamics as given by Eq.~\eqref{Hamilt} -- quadratic in momentum -- by a nonlinear dynamical system generically described by a Hamiltonian constraint, 
\begin{equation}
H^{W}(q,\,p) = K(p) + V(q),
\label{nlh}
\end{equation}
where evidently $\partial ^2 H^{W} / \partial q \partial p = 0$, and $K(p)$ and $V(q)$ are arbitrary functions of $p$ and $q$, respectively, recovers an equivalent Wigner continuity equation. It is cast in the form of Eq.~\eqref{alexquaz51}, with Wigner currents then given by
\begin{equation}
J_q(q,\,p;\,t) = +\sum_{\eta=0}^{\infty} \left(\frac{i\,\hbar}{2}\right)^{2\eta}\frac{1}{(2\eta+1)!} \, \left[\partial_p^{2\eta+1} K(p)\right]\,\partial_q^{2\eta}W(q,\,p;\,t),
\label{alexquaz500BB}
\end{equation}
and
\begin{equation}
J_p(q,\,p;\,t) = -\sum_{\eta=0}^{\infty} \left(\frac{i\,\hbar}{2}\right)^{2\eta}\frac{1}{(2\eta+1)!} \, \left[\partial_q^{2\eta+1} V(q)\right]\,\partial_p^{2\eta}W(q,\,p;\,t),\label{alexquaz500CC}
\end{equation}
which, also from Eq.~\eqref{alexquaz51}, leads to an explicit form of the stationarity quantifier given by \cite{Novo21A,Novo21B,Novo21C,Novo21D,Novo222}
\begin{equation} \label{helps}
\partial_t W= \sum_{\eta=0}^{\infty}\frac{(-1)^{\eta}\hbar^{2\eta}}{2^{2\eta}(2\eta+1)!} \, \left\{
\left[\partial_q^{2\eta+1}V(q)\right]\,\partial_p^{2\eta+1}W
-
\left[\partial_p^{2\eta+1}K(p)\right]\,\partial_q^{2\eta+1}W
\right\},\end{equation}
which encompasses all the contributions for quantum corrections of order $\mathcal{O}(\hbar^{2\eta})$, and to
a Liouvillianity quantifier (as from Eq.~\eqref{zeqnz59}) captured by \cite{Novo21A,Novo21B,Novo21C,Novo21D,Novo222}
\begin{equation}
\mbox{\boldmath $\nabla$}_{\xi} \cdot \mathbf{w} = \sum_{\eta=1}^{\infty}\frac{(-1)^{\eta}\hbar^{2\eta}}{2^{2\eta}(2\eta+1)!}%\times \,\nonumber\\&& 
\left\{
\left[\partial_p^{2\eta+1}K(p)\right]\,
\partial_q\left[\frac{1}{W}\partial_q^{2\eta}W\right]
-
\left[\partial_q^{2\eta+1}V(q)\right]\,
\partial_p\left[\frac{1}{W}\partial_p^{2\eta}W\right]
\right\}. ~~~\end{equation}
For Hamiltonians like those of the AAH form (cf. $H^{W}(q,\,p) = K(p) + V(q)$), the implementation of a {\em Hamiltonian function} through an {\em eigen}system, $H^{W}\, \psi_n = E_n\, \psi_n$ usually results into an unsolvable system.
Otherwise, the {\em Hamiltonian constraint} from Eq.~\eqref{nlh}, as opposed to a Hamiltonian function, allows for the identification of a probability flux continuity equation supported by the WW formalism, which can be helpful in discriminating classical from quantum behaviors, as well as phase-space equilibrium and stability properties, as it will be discussed in the following.

\section{Evaluation of the AAH system in the phase-space}

Keeping in mind the conditions for decoupling quantum corrections from nonlinear effects, before proceeding with the evaluation of the Hamiltonian dynamics from Eq.~\eqref{HamHarper01}, a more convenient perspective of the phase-space dynamics can be achieved if the Hamiltonian system is described in terms of a dimensionless version of $H^{W}(q,\,p)$ from Eq.~\eqref{nlh}, i.e. by \cite{Novo21A,Novo21B,Novo21C,Novo21D,Novo222}
\begin{equation}
\label{dimHH}\mathcal{H}(x,\,k) = \mathcal{K}(k) + \mathcal{V}(x),
\end{equation}
written in terms of dimensionless variables, $x = \left(m\,\omega\,\hbar^{-1}\right)^{1/2} q$ and $k = \left(m\,\omega\,\hbar\right)^{-1/2}p$.
In this case, one has $\mathcal{H} = (\hbar \omega)^{-1} H$, $\mathcal{V}(x) = (\hbar \omega)^{-1} V\left(\left(m\,\omega\,\hbar^{-1}\right)^{-1/2}x\right)$ and $\mathcal{K}(k) = (\hbar \omega)^{-1} K\left(\left(m\,\omega\,\hbar\right)^{1/2}k\right)$, where $m$ is a mass scale parameter and $\omega$ is an arbitrary angular frequency.
The Wigner function also cast into the dimensionless form of $\mathcal{W}(x, \, k;\,\omega t) \equiv \hbar\, W(q,\,p;\,t)$, suggests that $\hbar$ is absorbed by $dp\,dq\to \hbar\, dx\,dk$ integrations, with Wigner currents then written as $\mathcal{J}_x(x, \, k;\,\omega t)$ and $\mathcal{J}_k(x, \, k;\,\omega t)$, so as to have, $\omega\, \partial_x\mathcal{J}_x \equiv \hbar\, \partial_q J_q(q,\,p;\,t)$ and $\omega \,\partial_k\mathcal{J}_k\equiv \hbar \,\partial_p J_p(q,\,p;\,t)$, which finally can all be recast in the form of \cite{NossoPaper}
\begin{eqnarray}\label{alexDimW}
\label{imWA}\mathcal{J}_x(x, \, k;\,\tau) &=& +\sum_{\eta=0}^{\infty} \left(\frac{i}{2}\right)^{2\eta}\frac{1}{(2\eta+1)!} \, \left[\partial_k^{2\eta+1}\mathcal{K}(k)\right]\,\partial_x^{2\eta}\mathcal{W}(x, \, k;\,\tau),\\
\label{imWB}\mathcal{J}_k(x, \, k;\,\tau) &=& -\sum_{\eta=0}^{\infty} \left(\frac{i}{2}\right)^{2\eta}\frac{1}{(2\eta+1)!} \, \left[\partial_x^{2\eta+1}\mathcal{V}(x)\right]\,\partial_k^{2\eta}\mathcal{W}(x, \, k;\,\tau),
\end{eqnarray}
with
\begin{eqnarray}\label{alexDimW}
\mathcal{W}(x, \, k;\,\tau) &=& \pi^{-1} \int^{+\infty}_{-\infty} \hspace{-.35cm}dy\,\exp{\left(2\, i \, k \,y\right)}\,\phi(x - y;\,\tau)\,\phi^{\ast}(x + y;\,\tau),
\end{eqnarray}
where $y = \left(m\,\omega\,\hbar^{-1}\right)^{1/2} s$, $\tau = \omega t$,
such that the dimensionless continuity equation is written in terms of the re-defined phase-space coordinates, $\mbox{\boldmath $\xi$} = (x,\,k)$, as
\begin{equation}
{\partial_{\tau} \mathcal{W}} + {\partial_x \mathcal{J}_x}+{\partial_k \mathcal{J}_k} = {\partial_{\tau} \mathcal{W}} + \mbox{\boldmath $\nabla$}_{\xi}\cdot\mbox{\boldmath $\mathcal{J}$} =0.
\end{equation}

\subsection{Harper Hamiltonian and the AAH system classical correspondence}

Generic classes of Harper-like models, firstly introduced for parameterizing the behavior of electrons coupled to magnetic fields in a $2$-dim lattice \cite{Harper,HarperB}, turned into the departure framework for more complex systems. These encompass the description fractal structures related to the Hofstadter spectral decomposition \cite{Nat010,Nat020} and its connection with the phenomenology of the quantum Hall effect \cite{Prange}, as well as the identification of quantum mechanical topological phases, in particular, in the context of {\em designing} ultra-cold atom platforms for producing synthetic gauge fields and topological structures for neutral atoms \cite{001,002}.

From an effective perspective, Harper-like models can be reduced to a one-dimensional Hamiltonian formulation, capturing nearest-neighbor couplings with a sinusoidal modulation of the on-site energies. This is expressed through the following Hamiltonian constraint:
\begin{equation}\label{HamHarper00}
{H} \psi_n = -A_k (e^{+i\vartheta}\,\psi_{n+1} + e^{-i\vartheta}\,\psi_{n-1}) - A_x\,\cos(2\pi
\beta\, n +\theta) \psi_n,\end{equation}
where $A_{x}$ and $A_{k}$ represent the coupling strength and modulation amplitude, respectively \cite{PRL-Harper}. The phases $\theta$ and $\vartheta$ are associated with the wave vector in two dimensions.

In the above formulation, since the displacement of quantum numbers exhibited by $\psi_{n\pm1}$ in (\ref{HamHarper00}) corresponds to localized states in adjacent sites, quantum states depicted by $\psi_{n\pm1}\sim \psi({x\pm b}) \equiv \exp[\pm i\,k\,b]\psi(q)$ can be obtained from the translation operation, $\exp[\pm i\,k\,b]$, according to the application of the dimensionless momentum operator, $k$, for a coordinate correspondence given by $(x,\,b)\to(2\pi\beta n, 2\pi\beta)$.
The Hamiltonian Eq.~(\ref{HamHarper00}) then admits a semi-classical representation reduced to the form of sinusoidal contributions from Eq.~\eqref{HamHarper01} (i.e. with $w = 0$).
Hence, for corresponding coordinate operators, $\hat{x}$ and $\hat{k}$, satisfying $[\hat{x},\hat{k}] = i\,2\pi\beta$, the so-called Peierls phase parameter plays role of an effective Planck constant as it can be set by $2\pi\beta \equiv 1$ in the dimensionless analysis conduced in the previous sections, with $[\hat{x},\hat{k}] = i$.

\begin{figure}[h]
\includegraphics[scale=0.5]{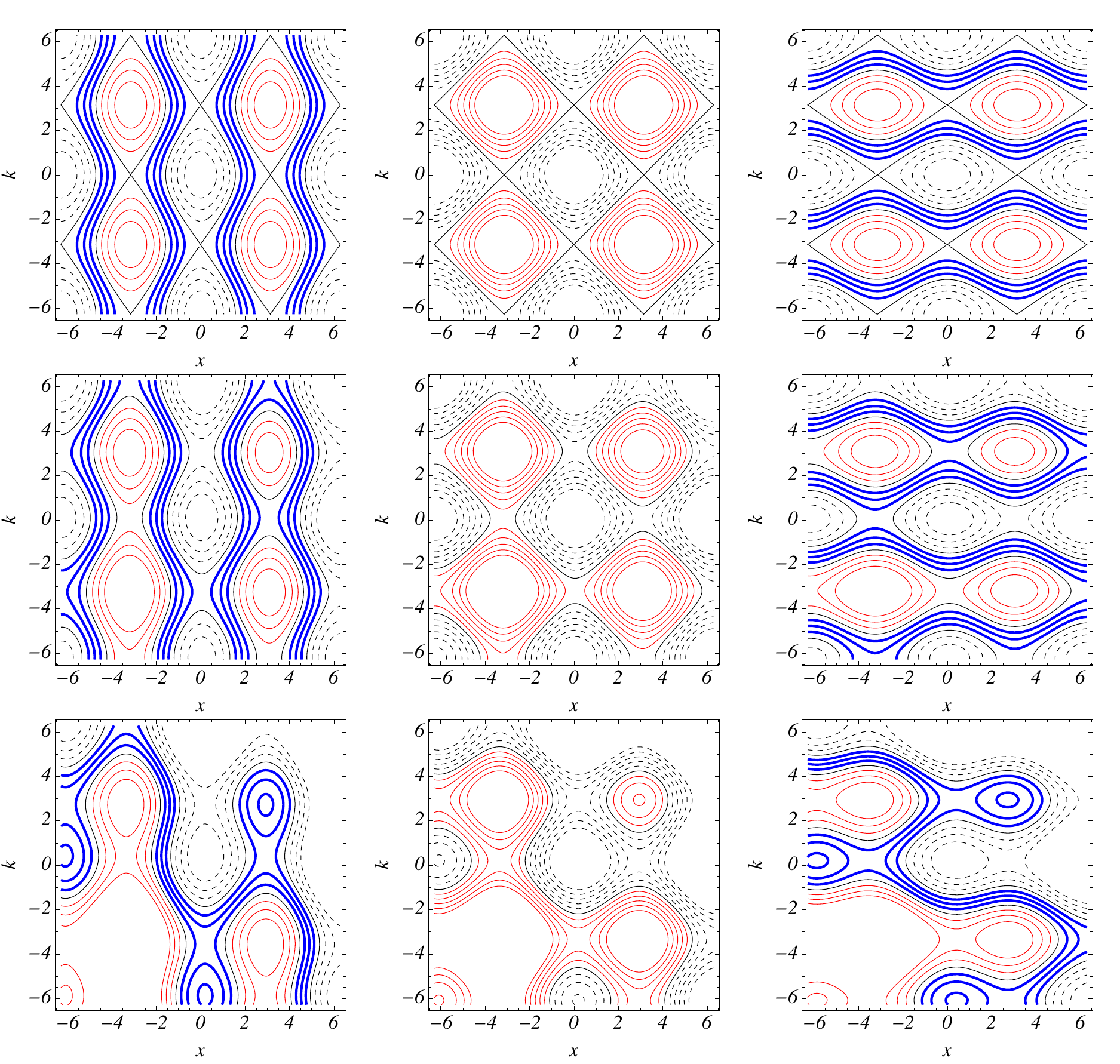}
\renewcommand{\baselinestretch}{.85}
\caption{\footnotesize
(Color online) Classical portrait of AAH Hamiltonians. Phase-space trajectories and corresponding lattice designs are for $ \mbox{Max}\{a^2-1,0\} < \vert\epsilon\vert < a^2+1$ corresponding to closed trajectories for $\epsilon > 0$ (black dashed lines) and for $\epsilon < 0$ (red thin lines), and for $0 < \vert\epsilon\vert < a^2-1$ corresponding to opened trajectories (blue thick lines), when they exist. All parameters are concerned with the original Harper pattern ($w=0$ (first row)). The limiar (opened-closed) value is given by $\vert\epsilon\vert = a^2 -1$. The plots are for $a^2=2$ (first column), with $\vert\epsilon\vert = 5/2,\, 2,\,3/2,\,\dots,\,0$, $a^2=1$ (second column), with $\vert\epsilon\vert = 5/2,\,2,\,3/2,\,\dots,\,0$ and $a^2=1/2$ (third column), with $\vert\epsilon\vert = 5/2,\, 2,\,3/2,\,\dots,\,0$. One also has $w=0$ (first row), $w=0.1$ (second row) and $w=0.4$ (third row), which have been chosen for comparative reasons.}
\label{HarperHarper}
\end{figure}

Furthermore, the phase-space representation of electron dynamics in a two-dimensional crystal is encapsulated by Eq.~(\ref{HamHarper00}). To account for nonlinear effects described by the Hamiltonian decomposition $\mathcal{H}(x, k) = \mathcal{K}(k) + \mathcal{V}(x)$, while disregarding the global sign and allowing for a phenomenological variation through an arbitrary parameter $a^2$, the system can be rewritten in the dimensionless form:
\begin{equation}\label{HamHarper01dim}
\mathcal{H}_H(x, k)= \cos(k) +a^2 \cos(x),
\end{equation}
which admits cyclic analytical solutions in a time-dependent framework.
The hypothesis of destroying the periodicity of phase-space trajectories is captured by the $H\to AAH$ Hamiltonian cast into the dimensionless form
\begin{equation}\label{HamHarper01dim}
\mathcal{H}_{_{AAH}}(x,\,k)= w\,k + \cos(k) +a^2(w\, x + \cos(x)),
\end{equation}
which exhibits (time-dependent) cyclic analytical solutions and therefore work as a feasible departure platform for identifying the implications related to the classicality and quantumness, namely, the Wigner flow deviations from stationary and Liouvillian regimes, according to the formalism introduced in the previous sections. The classical properties of the AAH Hamiltonian \eqref{HamHarper01dim} are depicted in Fig.~\ref{HarperHarper} where the phase-space trajectories associated to the corresponding lattice designs are identified for several values of the classical energy, $\mathcal{H}_{_{AAH}} \to \epsilon$, and for varying phenomenological parameters, $a$ and $w$. In particular, one has $w=0$ (first row), $w=0.1$ (second row) and $w=0.4$ (third row), which have been chosen for comparative reasons, from which the role of $w$ in destroying the periodicity of the classical system is evinced.

\subsection{Gaussian ensembles}

The framework discussed above can only be effective in quantifying quantum distortions from classical backgrounds if the Wigner currents are exactly computed.
Starting from Gaussian ensembles (cf. the Appendix I), stationarity and Liouvillianity quantifiers driven by $H^W(q,\,p)$ from Eq.~\eqref{nlh} can all be analytically computed through a procedure where the Wigner flow description results into a non-perturbative equivalent description of quantum fluctuations.

With the Gaussian distribution written as
\begin{equation}
\mathcal{G}_\alpha(x,\,k) = \hbar \,G_\alpha(q,\,p) = \frac{\alpha^2}{\pi}\, \exp\left[-\alpha^2\left(x^2+ k^2\right)\right],
\end{equation}
after some straighforward mathematical manipulations (cf. the Appendix I), one gets
\begin{eqnarray}
\label{imWA4}\partial_x\mathcal{J}^{\alpha}_x(x, \, k;\,\tau) &=& -2 \kappa(k)\,\sin\left(\alpha^2 \mu_{(k)}\,x\right)\,\exp[+\alpha^2 \mu^2_{(k)}/4]\,\mathcal{G}_{\alpha}(x, \, k)\,
,\\
\label{imWB4}\partial_k\mathcal{J}^{\alpha}_k(x, \, k;\,\tau) &=& +2 \upsilon(x)\,\sin\left(\alpha^2 \lambda_{(x)}\,k\right)\,\exp[+\alpha^2 \lambda^2_{(k)}/4]\,\mathcal{G}_{\alpha}(x, \, k),
\end{eqnarray}
which, as prescribed, points to a convergent series result for the stationarity quantifier, $\mbox{\boldmath $\nabla$}_{\xi}\cdot \mbox{\boldmath $\mathcal{J}$}^{\alpha}$, as well as, depending on the explicit form of the Hamiltonian, it can be manipulated in order to give the Liouvillian quantifier, $\omega^{-1}\mbox{\boldmath $\nabla$}_{\xi} \cdot \mathbf{w}$ and the complete pattern of the associated Wigner flow.

For Gaussian ensembles driven by $\mathcal{K}(k)=w\,k+\cos(k)$ and $\mathcal{V}(x)=a^2(w\,x + \cos(x))$, exact analytical results can be obtained.
From manipulations involving Eqs.~\eqref{t111} and \eqref{t222}, once they are replaced into Eqs.~\eqref{imWA4} and \eqref{imWB4}, one can write
\begin{eqnarray}
\label{imWA4CC}\partial_x\mathcal{J}^{\alpha}_x(x, \, k;\,\tau) &=& -2\left[w\,\alpha^2\,x - \,\sin\left(k\right)\,\sinh\left(\alpha^2\,x\right)\,\exp[-\alpha^2 /4]\right]\,\mathcal{G}_{\alpha}(x, \, k)\,
,\\
\label{imWB4CC}\partial_k\mathcal{J}^{\alpha}_k(x, \, k;\,\tau) &=& +2 a^2\left[w\,\alpha^2\,k - \sin\left(x\right)\,\sinh\left(\alpha^2\,k\right)\,\exp[-\alpha^2 /4]\right]\,\mathcal{G}_{\alpha}(x, \, k),
\end{eqnarray}
as a result from the convergent series expansion Eqs.~\eqref{imWA3} and \eqref{imWB3} in the Appendix I.

One can notice that, from the identified series expansion from Eqs.~\eqref{imWA3} and \eqref{imWB3} (as well as \eqref{imWA} and \eqref{imWB}), contributions from $\eta \geq 1$ introduce the quantum corrections which distort classical trajectories, i.e. quantum corrections follow from the coupling with the contributions due to the infinite expansion which are not suppressed by higher order derivatives at Eqs.~\eqref{imWA} and \eqref{imWB}. They emerge from nonlinear and non-quadratic Hamiltonian components which drive the nonlinear equations of motion in both classical and quantum pictures.

Finally, given that in the computation of the Wigner currents the quantum effects are accompanied by the nonlinear contributions from coordinate dependent potential and kinetic-like terms, $\mathcal{V}(x)$ and $\mathcal{K}(k)$, the exact result, Eqs.~\eqref{imWA4CC} and \eqref{imWB4CC}, obtained from the infinite expansion from Eqs.~\eqref{imWA3} and \eqref{imWB3} , guarantees that quantum corrections have been accurately accounted.
For the above scenario, the suppression of the $\eta \geq 1$ contributions results in a classical Hamiltonian dynamics yielded  by the classical Wigner currents as from Eqs.~\eqref{alexquaz500BB2} and \eqref{alexquaz500CC2}\footnote{In particular, by recasting Eqs.~\eqref{imWA4CC} and \eqref{imWB4CC} as:
\begin{eqnarray}
\label{imWA4CCD}\partial_x\mathcal{J}^{\alpha}_x(x, \, k;\,\tau) &=& \left[w - \,\sin\left(k\right)\,\frac{\sinh\left(\alpha^2\,x\right)}{\alpha^2\,x}\,\exp[-\alpha^2 /4]\right]\,\partial_x\mathcal{G}_{\alpha}(x, \, k)\,
,\\
\label{imWB4CCD}\partial_k\mathcal{J}^{\alpha}_k(x, \, k;\,\tau) &=& - a^2\left[w\ - \sin\left(x\right)\,\frac{\sinh\left(\alpha^2\,k\right)}{\alpha^2\,k}\,\exp[-\alpha^2 /4]\right]\,\partial_k\mathcal{G}_{\alpha}(x, \, k),
\end{eqnarray}
in the (Gaussian) delocalization limit of $\alpha\gtrsim 0$, one has
\begin{eqnarray}
\label{imWA4CCDD}\partial_x\mathcal{J}^{\alpha}_x(x, \, k;\,\tau) &\approx& \left[w - \,\sin\left(k\right)\,\right]\,\partial_x\mathcal{G}_{\alpha}(x, \, k) = +(\partial_k \mathcal{H})\,\partial_x\mathcal{G}_{\alpha}(x, \, k) 
,\\
\label{imWB4CCDD}\partial_k\mathcal{J}^{\alpha}_k(x, \, k;\,\tau) &\approx& - a^2\left[w\ - \sin\left(x\right)\right]\,\partial_k\mathcal{G}_{\alpha}(x, \, k) = -(\partial_x \mathcal{H})\,\partial_k\mathcal{G}_{\alpha}(x, \, k),
\end{eqnarray}
from which the Hamilton equations for classical statistical ensemble are identified (cf. Eqs.~\eqref{alexquaz500BB2} and \eqref{alexquaz500CC2}). This is equivalent to suppress the phase-space localization introduced by the Wigner distribution, which is coupled to the evolution of quantum contributions ($\eta \geq 1$).}

It means that a Gaussian ensemble driven by the AAH Hamiltonian can be considered in classical and quantum regimes, providing the exact analytic expression for the quantum distortions in the phase-space. 
The integrated Wigner currents obtained from Eqs.~\eqref{imWA4} and \eqref{imWB4} can thus be read as
\begin{eqnarray}
\label{imWA4CCDDD}\mathcal{J}^{\alpha}_x(x, \, k;\,\tau) &=& w\,\mathcal{G}_{\alpha}(x, \, k)\nonumber\\
&&+\frac{\alpha}{2\sqrt{\pi}} \,\sin\left(k\right)\,\exp\left(-\alpha^2\,k^2\right)\,
\left[\mbox{\sc{Erf}}\left(\alpha(x-1/2)\right)-\mbox{\sc{Erf}}\left(\alpha(x+1/2)\right)\right],\\
\label{imWB4CCDDD}\mathcal{J}^{\alpha}_k(x, \, k;\,\tau) &=& -a^2\bigg{\{}w\, \mathcal{G}_{\alpha}(x, \, k) \nonumber\\
&&\left.+\frac{\alpha}{2\sqrt{\pi}}\sin\left(x\right)\,\exp\left(-\alpha^2\,x^2\right)\,
\left[\mbox{\sc{Erf}}\left(\alpha(k-1/2)\right)-\mbox{\sc{Erf}}\left(\alpha(k+1/2)\right)\right]\right\},\qquad
\end{eqnarray}
written in terms of error functions, $\mbox{\sc{Erf}}(\dots)$, from which the components of the distorted quantum-like velocity, $\mathbf{w}$, $\omega_x$ and $\omega_k$, can be straightforwardly obtained as
\begin{eqnarray}\label{imWA4CC3mm}
\label{34}\omega_x &=& \mathcal{J}^{\alpha}_x(x,\,k;\,t)/\mathcal{G}_{\alpha}(x, \, k) \equiv f(x,\,k),\\
\label{35}\omega_k &=& \mathcal{J}^{\alpha}_x(x,\,k;\,t)/\mathcal{G}_{\alpha}(x, \, k) \equiv g(x,\,k).
\end{eqnarray}
\begin{figure}
\vspace{-1.2 cm}
\includegraphics[scale=0.23]{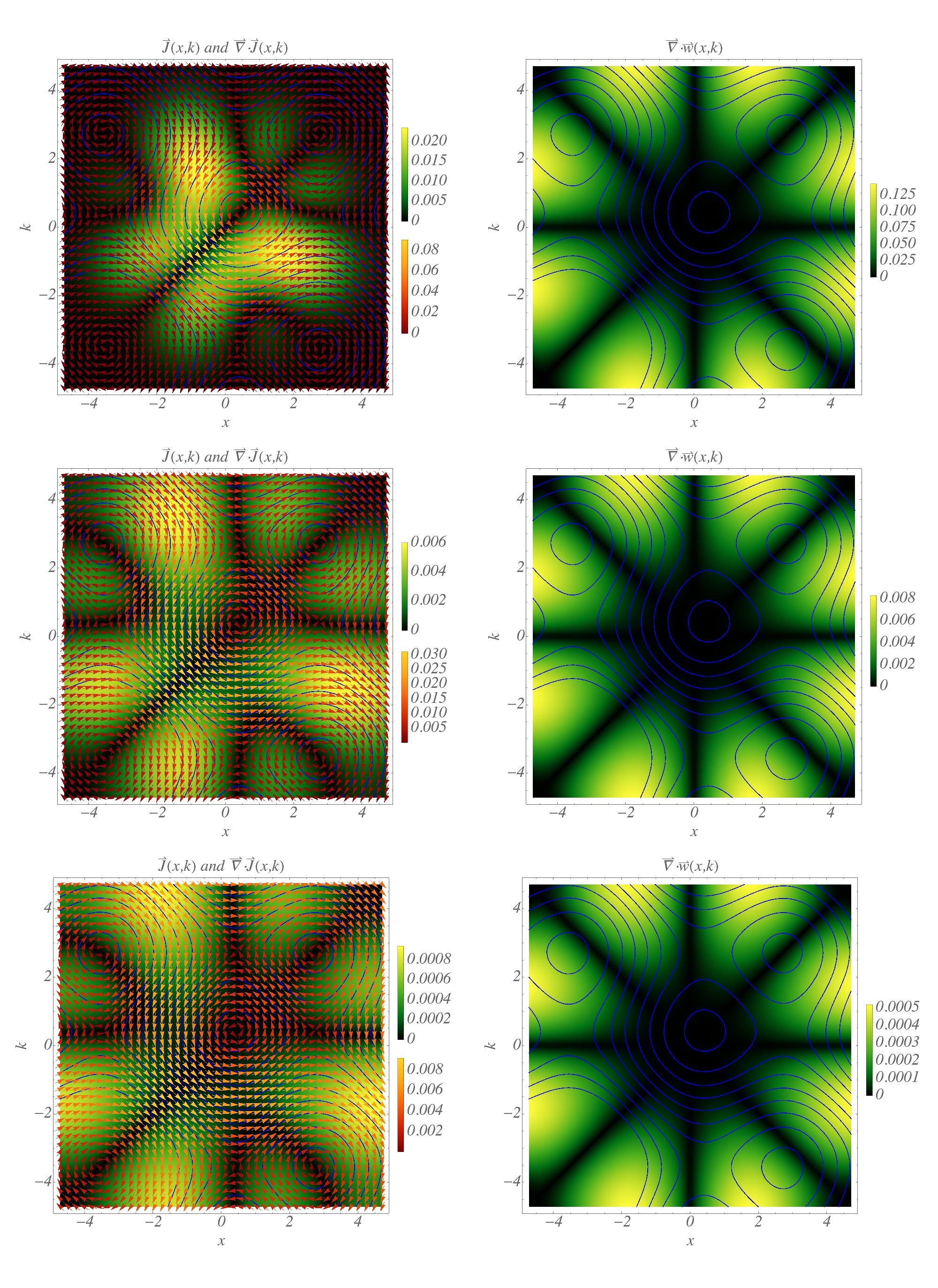}
\renewcommand{\baselinestretch}{.6}
\vspace{-1.2 cm}\caption{\footnotesize
(Color online)
{\em First column}: Features of the Wigner flow for the Gaussian ensemble, in the $x - k$ plane. At $\tau = 0$, Gaussian ensembles do not exhibit neither vortices nor stagnation points, in a kind of camouflage of the quantum distortions.
The stationarity quantifier, $\mbox{\boldmath $\nabla$}_{\xi} \cdot \mbox{\boldmath $\mathcal{J}$}^{\alpha}$, is described according to the background color scheme. The results are for the increasing spreading characteristic of the Gaussian function, from $\alpha =1/2$ (first row), $1/4$ (second row) and $1/8$ (third row). Peaked Gaussian distributions ($\alpha \gtrsim 1$) localizes the quantum distortions which result into non-stationarity. The parameter $w=0.4$ was arbitrarily chosen.
{\em Second column}: Liouvillian quantifier, $\omega^{-1}\mbox{\boldmath $\nabla$}_{\xi} \cdot \mathbf{w}$, depicted through the background color scheme, from darker regions, $\mbox{\boldmath $\nabla$}_{\xi} \cdot \mathbf{w} \sim 0$, to lighter regions, $\mbox{\boldmath $\nabla$}_{\xi} \cdot \mathbf{w} > 0$. 
In both columns, classical pattern is shown as a collection of black lines.
The stationarity quantifier, $\mbox{\boldmath $\nabla$}_{\xi} \cdot \mbox{\boldmath $\mathcal{J}$}^{\alpha}$, is described according to the background color scheme, from which lighter regions correspond to non-vanishing local contributions to $\partial_t \mathcal{G}_{\alpha}(x,\,k)$.}
\label{HarperHarper04}
\end{figure}
The Gaussian Wigner flow pattern described above is depicted in Fig.~\ref{HarperHarper04}, with the corresponding density plots for stationarity and Liouvillianity quantifieres, $\mbox{\boldmath $\nabla$}_{\xi} \cdot \mbox{\boldmath $\mathcal{J}$}^{\alpha}$ and $\omega^{-1}\mbox{\boldmath $\nabla$}_{\xi} \cdot \mathbf{w}$.	
Despite approaching classical-like closed orbits, now the perturbed quantum vortex of each periodic site (from $-\pi$ to $+\pi$) transverse surrounding values of $x$ and $k$ so as to mutually destroy the (classical) close orbit pattern.
The parameter $w=0.4$ was arbitrarily chosen for isotropic ($a=1$) $x-p$ planar patterns. Increasing values of $w$ completely destroys the equivalent crystal periodicity which, however, is only depicted by the Wigner currents (first column) since the contribution driven by $w$ is linear from the Hamiltonian, and therefore, it does not topologically modify the map from classical velocities to $\mathbf{w}$.

\section{Equilibrium points and hyperbolic stability}

\begin{figure}\label{GaussHarperAub}
\vspace{-1.2 cm}
\includegraphics[scale=0.46]{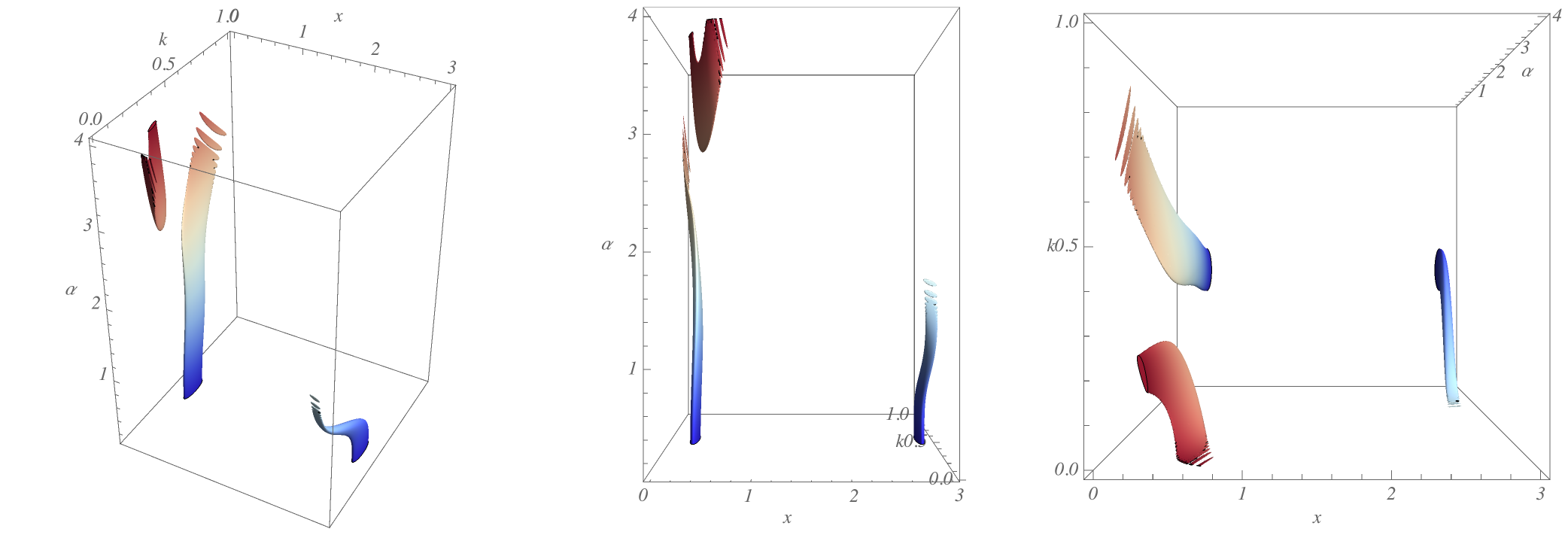}
\includegraphics[scale=0.46]{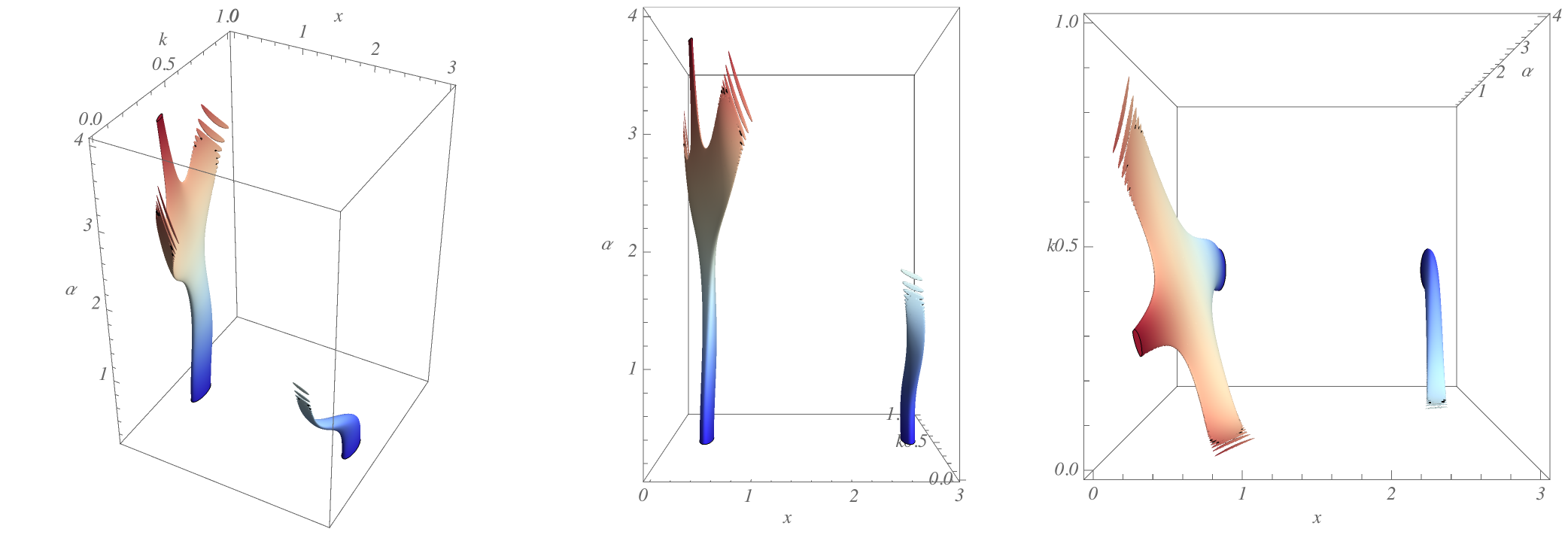}
\includegraphics[scale=0.46]{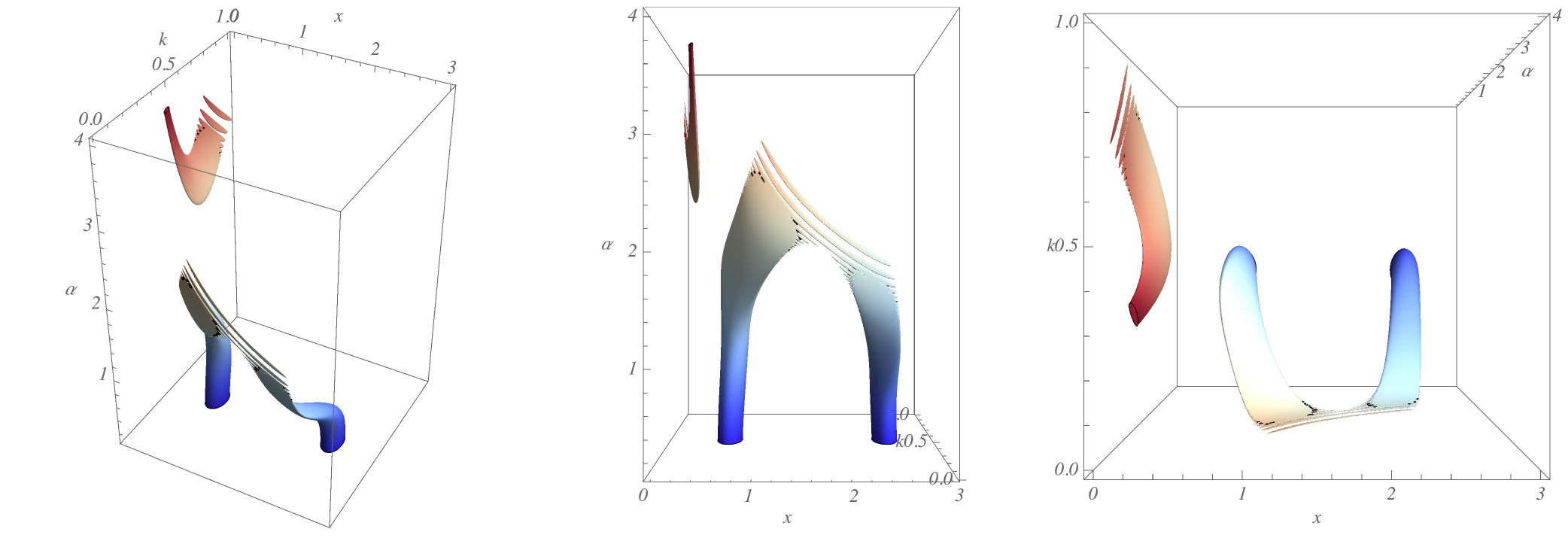}
\renewcommand{\baselinestretch}{.6}
\caption{\footnotesize
(Color online)
Region plot scheme for the phase-space evolution of quantum critical points corresponding to {\em quasi}-stable (blue regions) and unstable (red regions) equilibrium points in terms of the Gaussian spreading $\alpha$.
Results are for the Wigner flow with the equilibrium point (flux) surrounding envelop described by $\vert \vert\mathbf{w}\vert < 0.07$. For peaked Gaussians, $\alpha \gtrsim 3$, local effects compensate each other when sliced views of the Wigner flux for fixed $\alpha$ are considered, i.e. either when two vortices of opposite winding numbers match each other or when saddle points mutually annihilates one each other.
The spreading behavior of the Gaussian ensemble, from red bubble (unstable) islands to the blue (stable) envelop, corresponding to decreasing values of $\alpha$, diffusively recovers the classical-like pattern for which the quantum imprint is just to the small displacement of the ({\em quasi}-)stable equilibrium point. The results are for $w=4$ and $a=1.2$ (first row), $a=1$ (second row) and $a=0.8$ (third row). The portraits are the same for different angle views (columns).\label{Bio02}}
\end{figure}

From the above results, the Wigner flow pattern can be constrained by the form of $\mbox{\boldmath $\mathcal{J}$}^{\alpha} = \mathbf{w}\,\mathcal{G}_{\alpha}(x, \, k)$ such that the Gaussian flow pattern evaluated in terms of the parameter, $\alpha$, can now be interpreted in terms of the phase-space evolution of the equilibrium points, $\mathcal{J}^{{\alpha}}_x = \mathcal{J}^{{\alpha}}_k=0$.
The stagnation points, identified in the phase-space Wigner flow (cf. Fig.~\ref{GaussHarperAub}), can be evaluated in terms of the Gaussian parameter, $\alpha$, which drives the contributions due to the quantum fluctuations over the classical background. Fig.~\ref{Bio02} depicts the equilibrium point (flux) surrounding envelop with boundaries given by $\vert \vert\mathbf{w}\vert < 0.09$ for the same portraits viewed through different angles. The {\em quasi}-stable equilibrium point displacement is evinced by the blue region, for $\alpha \lesssim 3$. Despite approaching classical-like closed orbits, they are perturbed by a quantum vortex distortion which emerges from neighbor sites surrounding the values of $x$ and $k$ so as to break down the equilibrium point stability (darker red patterns). Such an analysis follows the same approach discussed for quantized eco-systems in previous works \cite{Novo21C,Novo21D,Novo222}.

For $\alpha > 4$ (highly peaked Gaussians), localizing appearance of unstable vortices and saddle points destroy the classical pattern. Red bubble regions correspond to topological phases which induce macroscopical modifications onto the periodic orbits. From red bubble saddle-point islands to the blue enveloped focus, decreasing values of $\alpha$, diffusively recovers the classical-like pattern ($\alpha = 0$).

\begin{figure}
\includegraphics[scale=0.375]{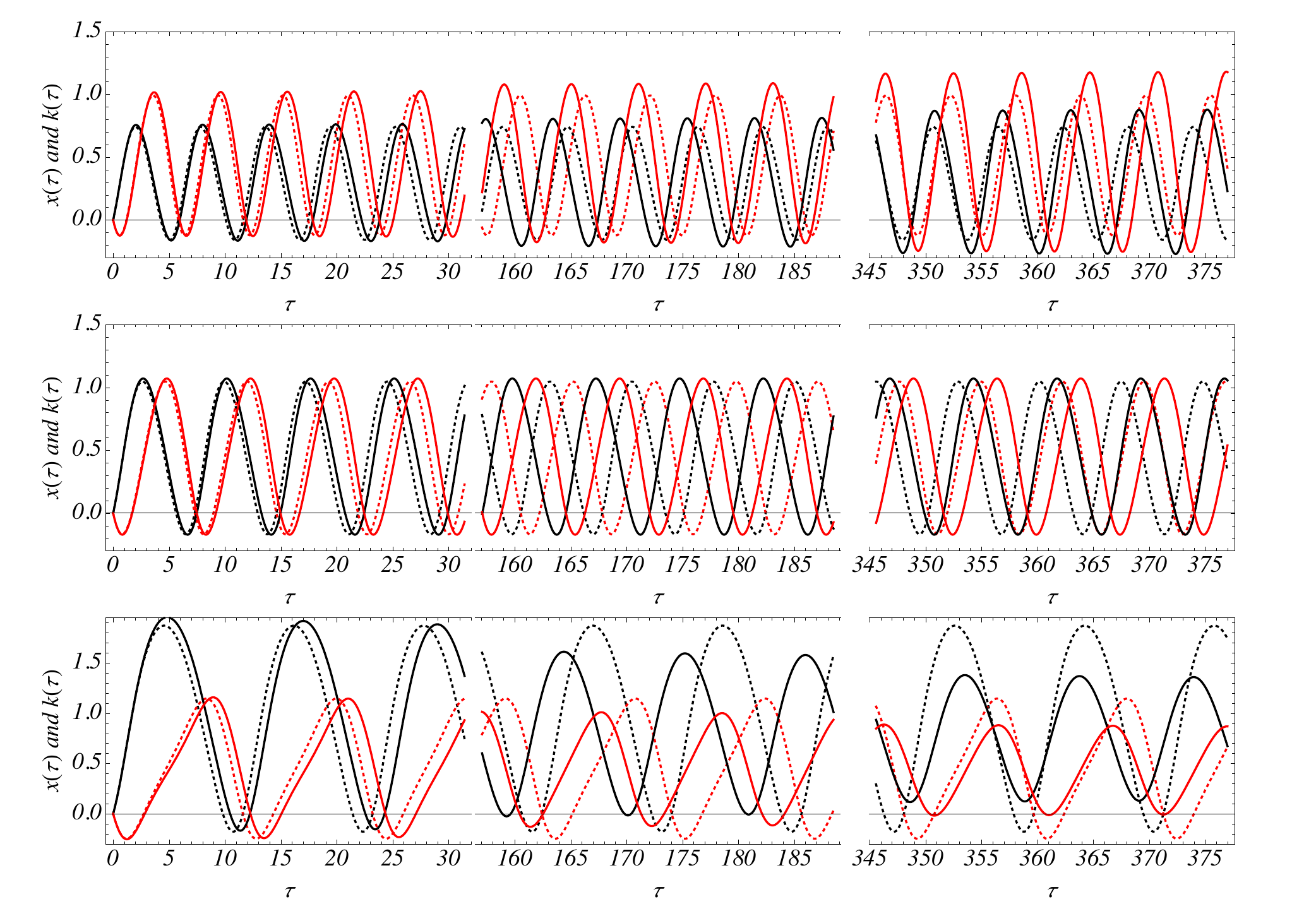}
\includegraphics[scale=0.247]{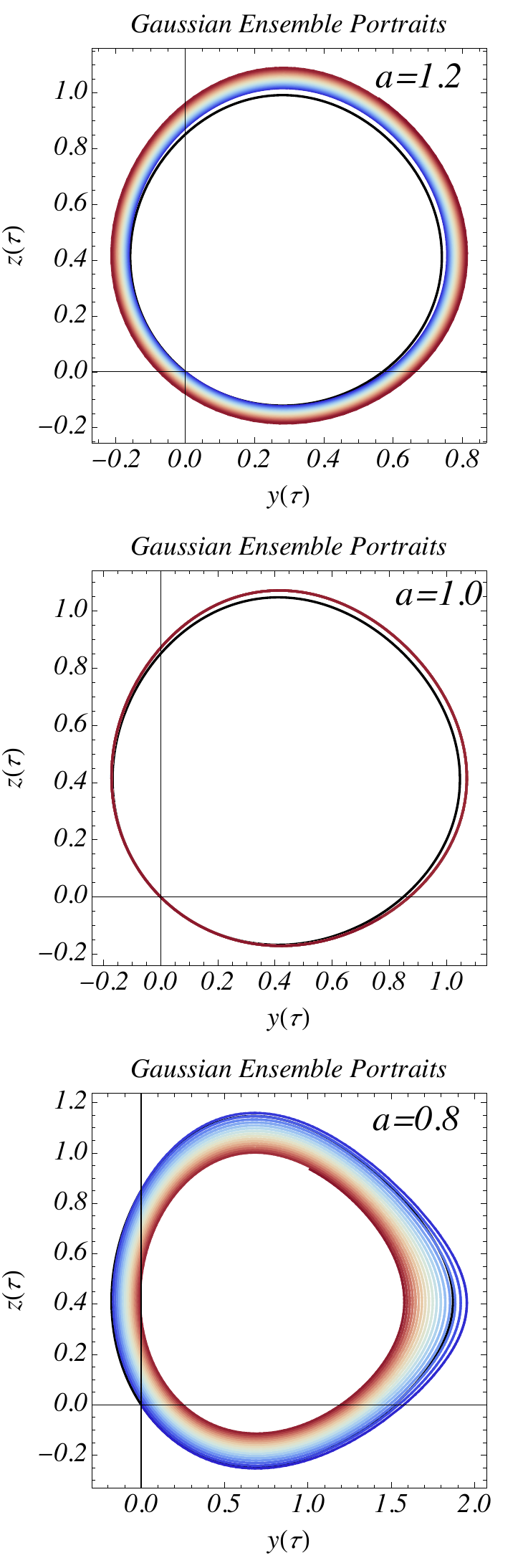}
\renewcommand{\baselinestretch}{.6}
\caption{\footnotesize
(Color online)
{\em First column}: Classical periodic (dashed lines) and quantum quasiperiodic (solid lines) dynamics, $x(\tau)$ (red lines) and $k(\tau)$, for typical spreading Gaussian ensembles, with $\alpha = 1/2$, $a = 1/2,\,1$ and $0.8$ and $w=0.4$.
{\em Second column}: Corresponding phase-space ({\em quasi}-)stable and quasiperiodic trajectories for classical (dashed lines) and quantum (solid lines) patterns. The color scheme describes the quantum {\em quasi}-stable evolution from $\tau = 0$ (blue tone) to $\tau \gg 0$ (red tone).
.}
\label{Bio03}
\end{figure}

For visualizing the quasiperiodicity related to the quantum imprints, one can follow the effective approach which accounts for the distortions obtained from Eqs.~\eqref{imWA4CC3mm}.
Through the quantum analogue phase-space velocity components, one can numerically compute $x(\tau)$ and $k(\tau)$ so as to visualize how the AAH trajectories are macroscopically affected by the equilibrium point averaged quantum displacement at almost stable regimes. The results are depicted in Fig.~\ref{Bio03} for typical spreading Gaussian ensembles, with $a = 1.2,\,1$ and $0.8$, $w = 0.4$, and with $\alpha = 1/2$.
The deviations provide by $a\neq 1$ just indicate unstable ($a > 1$) and stable ($a < 1$) quasiperiodic trajectories.
For $a = 1$, a stable equilibrium point configuration is quickly restored, which can be numerically verified for any Gaussian configuration with $\alpha \lesssim 2$.

Finally, in the context of the classical phase-space dynamics, results from Fig.~\ref{Bio03} correspond to the solutions of the system of ordinary differential equations (cf. Eq.~\eqref{imWA4CC3mm}) with stationary behavior. 
Therefore, the equilibrium is geometrically defined by $\dot{\mbox{\boldmath{$\xi$}}} = 0$ (i.e. $v_{x(\mathcal{C})}=v_{k(\mathcal{C})}=0$), which has a straightforward quantum correspondence expressed in terms of the quantum velocity, $\mathbf{w}$, by $\omega_x=\omega_k=0$ from Eq.~\eqref{imWA4CC3mm}. In both classical and quantum descriptions, according to the elementary theoretical grounds for describing the so-called hyperbolic stability (cf. the Appendix II), equilibrium points correspond to the phase-space stagnation points. 

Since the hyperbolic equilibrium admits small linear perturbations over the system of equations, the phase-space portrait qualitatively does not deviate from the equilibrium configuration.
Hence, the local phase portrait of a nonlinear system can be mapped by its linearized version which equivalently accounts for eventual short displacements of the fixed points (cf. the Hartman-Grobman theorem \cite{HG}).
Conversely, several types of non-hyperbolic equilibrium patterns result into local bifurcations which may change stability, suppress the fixed point features, or even split them into several equilibrium points, as it is qualitatively depicted in Fig.~\ref{Bio03}.
According to the results shown in Fig.~\ref{Bio03}, cf. Eqs.~(\ref{34}) and (\ref{35}) (and Eqs.~(\ref{imWA4CCD}) and (\ref{imWB4CCD})), the equilibrium point classifications are englobed by the hyperbolic equilibrium criterium only for $\alpha \lesssim 3$, where short displacements from classical configurations ($\alpha=0$) are evinced by the blue region. 

\begin{figure}
\vspace{-1.2 cm}
\includegraphics[scale=0.62]{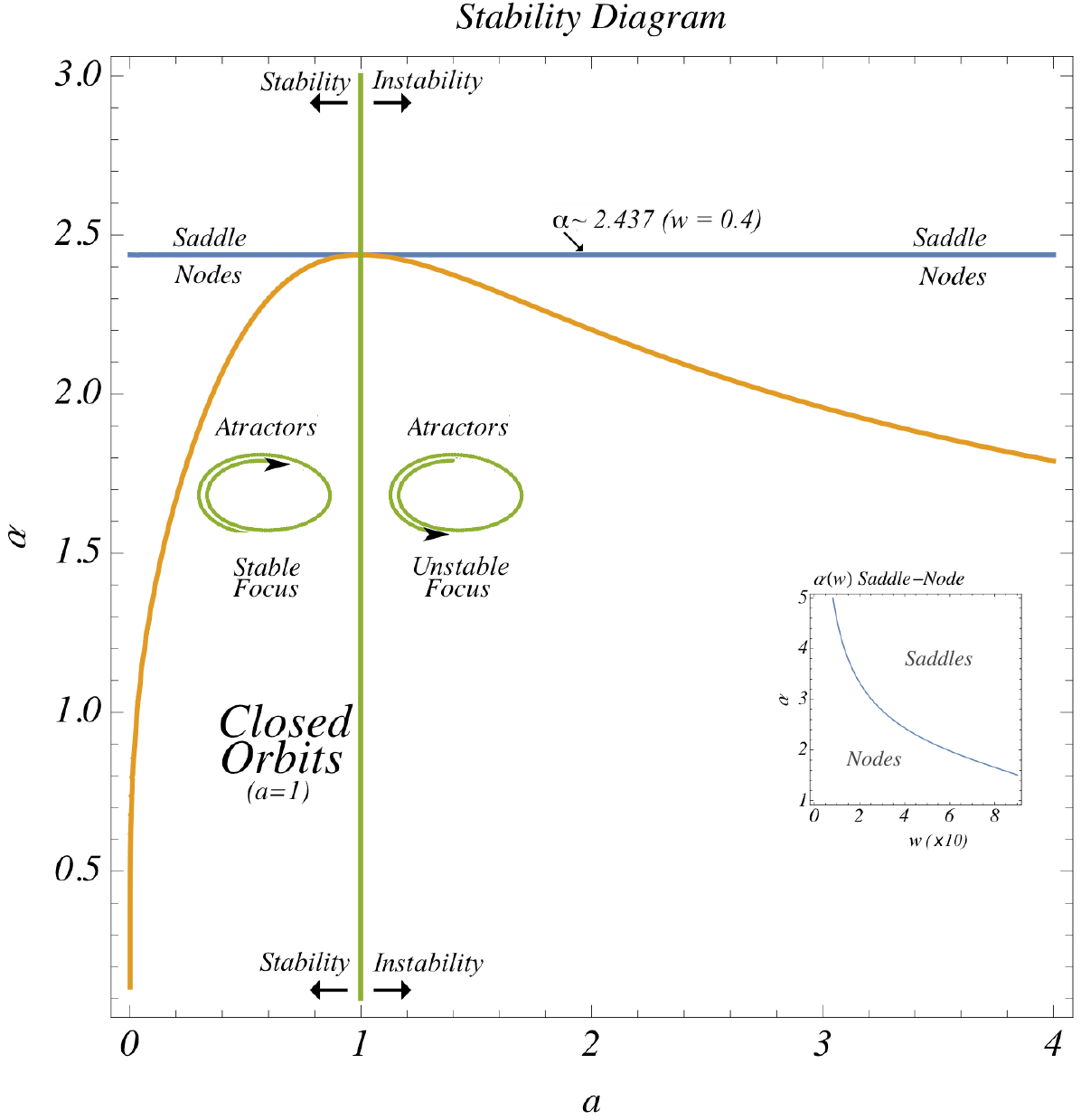}
\renewcommand{\baselinestretch}{.6}
\caption{\footnotesize
(Color online)
Hyperbolic equilibrium and stability for the AAH system, obtained from Eqs.~(\ref{imWA4CCD4mm})-(\ref{imWABCCD4mm}) as function of the $\alpha(w)$ and the anisotropy parameter $a$.
\label{Bio0004}
}
\end{figure}

The complete hyperbolic equilibrium pattern can be summarized by the diagrams from Fig.~\ref{Bio0004}, were phase-space saddle points correspond to $\alpha \gtrsim 2.437$, from $Det[j(x_o,\,k_o)] < 0$, in opposition to focus and node points identified by $\alpha \lesssim 2.437$, from $Det[j(x_o,\,k_o)] > 0$. The attractor regimes, for $\Delta[j(x_o,\,k_o)]>0$, are defined in terms of the anisotropy factor, $a$, which also sets the threshold for stability at $a=1$.
Results are for $w = 0.4$, but the corresponding dependence of the equilibrium threshold value for $\alpha$ as function of $w$ is numerically described in the subplot of Fig.~\ref{Bio0004}.

For increasing values of $\alpha$, the saddle-points which emerge from the lighter white patterns shown in Fig.~\ref{Bio03} naturally contributes to the subsequent diffusive appearance of unstable vortices and additional saddle-points that completely erase the classical pattern where the red bubble regions correspond to the quantum drivers for instabilities.
However, in this case, the hyperbolic equilibrium classification cannot be consistently used since the system instabilities do not support a perturbative linear approximation.

To clear up this point, the impact of quantum corrections can also be evaluated in terms of the semi-classical analysis of the so-called hyperbolic equilibrium and stability conditions, described in terms of the variables $x(\tau)$ and $k(\tau)$. Turning back to Eqs.~\eqref{imWA4CCDDD} and \eqref{imWB4CCDDD} (and Eqs.~\eqref{34} and \eqref{35}) and expanding them up to order $\mathcal{O}(\alpha^4)$, associated quantum-analogue velocities can be recast in terms of the time derivatives of $x(\tau)$ and $k(\tau)$, i.e.
\begin{eqnarray}
\label{imWA4CCDDDm}\frac{d{x}}{d\tau} &\equiv& 
\frac{\mathcal{J}^{\alpha}_x}{\mathcal{G}_{\alpha}} = \omega_x = f(x, \,k)\nonumber\\
&=& w +\frac{\sqrt{\pi}}{2\alpha} \,\sin\left(x\right)\,\exp\left(\alpha^2\,x^2\right)\,
\left[\mbox{\sc{Erf}}\left(\alpha(x-1/2)\right)-\mbox{\sc{Erf}}\left(\alpha(x+1/2)\right)\right]\nonumber\\
&\approx& w - \sin(k) \left[1-\frac{\alpha^2}{12}+\frac{\alpha^4}{480}(3+80x^2)\right],\\
\label{imWB4CCDDDm}\frac{d{k}}{d\tau}&\equiv& 
\frac{\mathcal{J}^{\alpha}_k}{\mathcal{G}_{\alpha}} = \omega_k = g(x, \,k)\nonumber\\
&=& -a^2\left\{w +\frac{\sqrt{\pi}}{2\alpha}\sin\left(k\right)\,\exp\left(\alpha^2\,k^2\right)\,
\left[\mbox{\sc{Erf}}\left(\alpha(k-1/2)\right)-\mbox{\sc{Erf}}\left(\alpha(k+1/2)\right)\right]\right\}\nonumber\\
&\approx& -a^2\left\{w - \sin(x) \left[1-\frac{\alpha^2}{12}+\frac{\alpha^4}{480}(3+80k^2)\right]\right\},\qquad
\end{eqnarray}

From these quantum-modified equations, a further stability analysis of the equilibrium points in terms of the Gaussian localization parameter, $\alpha$, can be performed through the Jacobian matrix, $j(x,\,k)$ (cf. Eq.~\eqref{imWA4CCD4mm} in the Appendix II)  evaluated at the equilibrium points obtained from $$\frac{d{x}}{d\tau}\bigg{\vert}_{eq} = \frac{d{k}}{d\tau}\bigg{\vert}_{eq} =f(x_o, \,k_o)=g(x_o,\,k_o)=0,$$
with $f(x, \,k)$ and $g(x,\,k)$ identified from Eqs.~\eqref{imWA4CCDDDm} and Eqs.~\eqref{imWB4CCDDDm}, respectively.
Second-order contributions from $\alpha^2$ merely shift the equilibrium points from $x_o = k_o = \arcsin{(w)}$ to $x_o = k_o \approx \arcsin{[w(1-\alpha^2/12)^{-1}]}$ and do not affect the stability conditions of the closed orbits in phase-space, which are characterized by $Tr[j(x_o, \,k_o)] \approx 0$ with $Det[j(x_o, \,k_o)] = a^2(1-w^2-\alpha^2/6)$\footnote{According to the hyperbolic stability criterion, the stability properties of focus and node points are determined by the trace as
\begin{eqnarray} Tr[j(x_o, \,k_o)] > 0 \qquad &\to& \mbox{instability},\nonumber\\ Tr[j(x_o, \,k_o)] < 0 \qquad &\to& \mbox{stability},\nonumber \end{eqnarray}
for $Det[j(x_o, \,k_o)] > 0$.
Saddle points arise when $Det[j(x_o, \,k_o)] < 0$. Therefore, only for higher values of $\alpha$ the transition from nodes to saddle points are admitted (cf. the numerically obtained limiar for $\alpha$ and $w$ correspondence depicted the small rectangle in the plot of Fig.~\ref{Bio0004}.}.

Considering iterative corrections of order $\mathcal{O}(\alpha^4)$, one finds
\begin{eqnarray} Tr[j(x_o, \,k_o)] \approx\frac{\alpha^4}{3} (a^2-1) \,w\,\arcsin(w), \end{eqnarray}
from which, considering the equilibrium regime with  $Det[j(x_o, \,k_o)]>0$, it follows that $\mathcal{O}(\alpha^4)$ corrections drive the system towards stable ($a < 1$) and unstable ($a > 1$) domains, as depicted in Fig.~\ref{Bio03}\footnote{The plots in Fig.\ref{Bio03} are obtained by numerically solving the exact expressions for the time derivatives of $x(\tau)$ and $k(\tau)$, derived from Eqs.~\eqref{imWA4CCDDD} and \eqref{imWB4CCDDD}.}.

Interpreting these phase-space oscillatory dynamics as the {\it quantum distortion} of the system one can state that, for $a > 1$ and shortly increasing values of $\alpha$, with $\alpha \gtrsim 0$, quantum effects asymptotically drive the system towards  quasi-periodic cycles with expanding amplitudes corresponding to an unstable focus.
Similarly, for $a < 1$ and $\alpha \gtrsim 0$, oscillations are asymptotically suppressed, leading to stabilization around the modified equilibrium points $x_o = k_o \approx \arcsin{[w(1-\alpha^2/12)^{-1}]}$.

Therefore, the {\it quantum analog} hypothesis of the AAH dynamics discussed here, where either $[x,,k] = i$ or, more generally, $[x,\,k] \neq 0$, reveals that quantum effects, localized by Gaussian ensembles, with localization driven by increasing values of $\alpha$, lead to an explicit reformulation of the equations of motion, suggesting a (quantum) modified dynamics.

\section{Conclusions} 

The hyperbolic equilibrium configurations for the AAH dynamics have been analytically evaluated according to the Wigner flow framework for quantifying quantum fluctuations over a classical background. A consistent map of hyperbolic stability conditions related to nonlinear Hamiltonians in the form of $H^{W}(q,\,p) = K(p) + V(q)$, was presented, from which quantum fluctuations over the equivalent classical background were examined. As discussed in previous works, the extended version of the WW framework for nonlinear Hamiltonian systems is a suitable probe for both quantumness and classicality in the context of AAH systems. The combination of the hyperbolic stability quantifiers with the Wigner features was shown to be relevant in distinguishing quantum fluctuations from nonlinear effects, particularly when limitations of the Schr\"odinger framework are identified.

Extrapolating from the hyperbolic equilibrium framework to the study of deterministic chaos, for continuous systems, it is also associated with the nonlinearity in Hamiltonian systems. Changes to the topological structure, sometimes expressed by local bifurcations, occur when a parameter change affects the stability of an equilibrium (or fixed point). This is driven by the real part of an eigenvalue of an equilibrium pattern passing through zero, resulting in a bifurcation point where the equilibrium becomes non-hyperbolic. Considering the results obtained here, no significant deviations from the hyperbolic domain were found.

More specifically in the context of AAH-modulated systems, the role of nonlinear effects has been suitably addressed in recent investigations. AAH-modulated system properties have been identified in various experimental platforms, including nonlinear photonic lattices, ultracold atomic gases, nonlinear quantum walks, and non-Hermitian systems. In these systems, nonlinearities significantly influence localization, transport, and topological properties, emphasizing their influence in shaping the behavior of quasiperiodic systems.
In particular, quasiperiodic photonic lattices with Kerr-type nonlinearity have been shown to support solitonic states and breathers, where the balance between nonlinearity and localization leads to nontrivial transport and topological phase transitions \cite{Lahini,Segev2013,Smirnova2020}. Likewise, Bose-Einstein condensates trapped in optical lattices with AAH-type quasiperiodicity exhibit interaction-induced self-trapping and mobility edges, where nonlinear effects arising from atom-atom interactions modify the localization properties \cite{Modugno2009,Roati,Schreiber2015}.
More recently, discrete-time quantum walks with AAH-type quasiperiodicity and nonlinear feedback mechanisms demonstrate dynamically controlled localization-delocalization transitions, where nonlinearity directly influences the spreading properties of the wavefunction \cite{Rakovszky2017,Cedzich2021}.
Even in the context of non-Hermitian quantum mechanics, the interplay between gain-loss nonlinearity and AAH modulation has been investigated in PT-symmetric optical lattices, leading to phase transitions sensitive to nonlinear effects \cite{Longhi2019}. 
And as it was previously mentioned, periodically driven AAH models have also been explored in the context of Floquet engineering, where nonlinearities give rise to emergent topological states \cite{Zhou2021}.

All these experimental realizations suggest that nonlinear effects are not only observable but also play a critical role in modifying localization, transport, and topological properties in AAH-modulated systems. Hence, the framework for the  quantum modified (theoretical) scenario here addressed also corroborates with the above-mentioned interplay between nonlinearity and quasiperiodicity, and therefore deserves continuous investigations.

As a summarizing remark, it is noteworthy that the application of the phase-space WW formulation for discussing equilibrium and stability of nonlinear Hamiltonian systems, as described by AAH systems, can be extended to several other quasicrystal systems. These include the one-dimensional tight-binding model with periodic diagonal incommensurate potentials (the so-called Maryland model \cite{Grempel}), as well as derived approaches for describing the fractal spectrum and anomalous diffusion behavior in the Harper model \cite{Shape,Artuso01,Artuso02,Zhao}, which might be connected with exactly solvable almost-periodic Schr\"odinger operators \cite{Simon}. All these platforms admit experimental realizations and, therefore, are suitable for discussing non-equilibrium and phase-transition phenomena.
In such a broader context, it is worth mentioning that the framework discussed here can be extended to other more complex forms of nonlinear Hamiltonians, where chaotic patterns and quantum fluctuations \cite{Novo21A, Novo21B, Novo21C, Novo21D, Novo222} could coexist.

Finally, our results confirm that a broad class of nonlinear Hamiltonian systems, subjected to quantum mechanical and equilibrium paradigms, exhibit regular configurations that, in the context of the WW phase-space framework, can be detected either microscopically through quantum topological phase indirect observations or macroscopically through time-evolved averaged-out statistical imprints. These possibilities reinforce the growing interest in the extended framework of phase-space quantum mechanics and suggest its potential for further applications in nonlinear systems exhibiting quantum and classical phenomena.

{\em Acknowledgments -- The work of AEB is supported by the Brazilian Agencies FAPESP (Grant No. 2023/00392-8, S\ ~ao Paulo Research Foundation (FAPESP)) and CNPq (Grant No. 301485/2022-4).}

\section*{Appendix I -- Evaluation of Gaussian ensembles}

For a Gaussian distribution written as
\begin{equation}
\mathcal{G}_\alpha(x,\,k) = \hbar \,G_\alpha(q,\,p) = \frac{\alpha^2}{\pi}\, \exp\left[-\alpha^2\left(x^2+ k^2\right)\right],
\end{equation}
the associated Wigner flow contributions assume the form of
\begin{eqnarray}
\label{imWA2}\partial_x\mathcal{J}_x(x, \, k;\,\tau) &=& +\sum_{\eta=0}^{\infty} \left(\frac{i}{2}\right)^{2\eta}\frac{1}{(2\eta+1)!} \, \left[\partial_k^{2\eta+1}\mathcal{K}(k)\right]\,\partial_x^{2\eta+1}\mathcal{G}_{\alpha}(x, \, k),
\\
\label{imWB2}\partial_k\mathcal{J}_k(x, \, k;\,\tau) &=& -\sum_{\eta=0}^{\infty} \left(\frac{i}{2}\right)^{2\eta}\frac{1}{(2\eta+1)!} \, \left[\partial_x^{2\eta+1}\mathcal{V}(x)\right]\,\partial_k^{2\eta+1}\mathcal{G}_{\alpha}(x, \, k),
\end{eqnarray}
once the Hamiltonian Eq.~(\ref{dimHH}) is considered.
From Gaussian relations with Hermite polynomials of order $n$, $\mbox{\sc{h}}_n$, one has
\begin{equation}
\partial_\zeta^{2\eta+1}\mathcal{G}_{\alpha}(x, \, k) = (-1)^{2\eta+1}\alpha^{2\eta+1}\,\mbox{\sc{h}}_{2\eta+1} (\alpha \zeta)\, \mathcal{G}_{\alpha}(x, \, k),
\end{equation}
for $\zeta = x,\, k$, which can be reintroduced into Eqs.~\eqref{imWA2} and \eqref{imWB2} as to lead to potentially convergent series expansions. This allows for recasting the Wigner flow expressions in an analytical form, which accounts for the overall quantum distortion contributions, i.e. for $\eta$ from $0$ to $\infty$ into Eqs.~(\ref{imWA2})-(\ref{imWB2}).
In particular, for the quantum systems where $\mathcal{V}$ and $\mathcal{K}$ derivatives can be recast in the form of
\begin{eqnarray}
\label{t111}
\partial_x^{2\eta+1}\mathcal{V}(x) &=& \lambda^{2\eta+1}_{(x)} \, \upsilon(x),\\
\label{t222}
\partial_k^{2\eta+1}\mathcal{K}(k) &=& \mu^{2\eta+1}_{(k)} \, \kappa(k),
\end{eqnarray}
with $\lambda$, $\upsilon$, $\mu$, and $\kappa$ being arbitrary auxiliary functions, it can be straightforwardly verified that, once substituted into Eqs.~\eqref{imWA2} and \eqref{imWB2}, the above expressions lead to
\begin{eqnarray}
\label{imWA3}\partial_x\mathcal{J}_x(x, \, k;\,\tau) &=& (+2i) \kappa(k)\,\mathcal{G}_{\alpha}(x, \, k)\,\sum_{\eta=0}^{\infty} \left(\frac{i\,\alpha\,\mu_{(k)}}{2}\right)^{2\eta+1}\frac{1}{(2\eta+1)!} \, \mbox{\sc{h}}_{2\eta+1} (\alpha x),\\
\label{imWB3}\partial_k\mathcal{J}_k(x, \, k;\,\tau) &=& (-2i) \upsilon(x)\,\mathcal{G}_{\alpha}(x, \, k)\sum_{\eta=0}^{\infty} \left(\frac{i\,\alpha\, \lambda_{(x)}}{2}\right)^{2\eta+1}\frac{1}{(2\eta+1)!} \, \mbox{\sc{h}}_{2\eta+1} (\alpha k).\end{eqnarray}
Finally, by noticing that
\begin{equation}
\sum_{\eta=0}^{\infty}\mbox{\sc{h}}_{2\eta+1} (\alpha \zeta)\frac{s^{2\eta+1}}{(2\eta+1)!} = \sinh(2s\,\alpha\zeta) \exp[-s^2],
\end{equation}
one gets Eqs.~\eqref{imWA4} and \eqref{imWB4}.

\section*{Appendix II -- Hyperbolic equilibrium and stability conditions}

From features of the Jacobian matrix,
\begin{equation}\label{imWA4CCD4mm}
j (x,\,k) = \left[\begin{array}{rr}
\partial_x f(x,\,k) & \partial_k f(x,\,k)\\
\partial_x g(x,\,k) & \partial_k g(x,\,k)
\end{array}\right],
\end{equation}
an approximated criterium for linear stability can be obtained from its eigenvalues.
Hyperbolic equilibrium and stability conditions are stratified into subclassifications, through the trace, ${Tr}[\dots]$, and the determinant, ${Det}[\dots]$, of $j (x,\,k)$, when all derivatives are evaluated at the equilibrium point, ${\mbox{\boldmath{$\xi$}}_o}= (x_o,\,k_o)$, i.e. from $f(x_o,\,k_o)=g(x_o,\,k_o)=0$.

One has $j(x_o,\,k_o)$ with all the eigenvalues with negative real parts for asymptotically stable systems and least one eigenvalue with a positive real part for unstable systems.
The Jacobian matrix drives the conditions for the so-called hyperbolic equilibrium if all their eigenvalues have non-zero real parts and, if at least one of its eigenvalue at equilibrium points has a zero real part, then the equilibrium is not hyperbolic. In this last case, the robustness of equilibrium and stability conditions imposes an enhanced classification \cite{Book,Book2}.

From the above criteria, focus and node stabilities are defined by trace properties as 
\begin{eqnarray}
Tr[j(x_o,\,k_o)] &=& \mbox{\boldmath $\nabla$}_{\xi}\cdot\mathbf{w}\vert_{{\mbox{\boldmath{$\xi$}}_o}} > 0 \qquad \to \mbox{instability},\nonumber\\
Tr[j(x_o,\,k_o)] &=& \mbox{\boldmath $\nabla$}_{\xi}\cdot\mathbf{w}\vert_{{\mbox{\boldmath{$\xi$}}_o}} < 0 \qquad \to \mbox{stability},
\end{eqnarray}
when 
\begin{eqnarray}
Det[j(x_o,\,k_o)] &=& \partial_x f\vert_{{\mbox{\boldmath{$\xi$}}_o}}\,\partial_k g\vert_{{\mbox{\boldmath{$\xi$}}_o}}- \partial_k f\vert_{{\mbox{\boldmath{$\xi$}}_o}}\,\partial_x g\vert_{{\mbox{\boldmath{$\xi$}}_o}} > 0 \qquad \to \mbox{for focus and nodes},
\end{eqnarray}
and one has saddle points for
\begin{eqnarray}
Det[j(x_o,\,k_o)] &=& \partial_x f\vert_{{\mbox{\boldmath{$\xi$}}_o}}\,\partial_k g\vert_{{\mbox{\boldmath{$\xi$}}_o}}- \partial_k f\vert_{{\mbox{\boldmath{$\xi$}}_o}}\,\partial_x g\vert_{{\mbox{\boldmath{$\xi$}}_o}} < 0 \qquad \to \mbox{for saddle points}.
\end{eqnarray}
Focus and nodes are also separated by $\Delta[j] = Tr[j]^2 - 4 Det[j] = 0$ as
\begin{eqnarray}\label{imWABCCD4mm}
\Delta[j(x_o,\,k_o)] && > 0 \qquad \mbox{for nodes},\nonumber\\
\Delta[j(x_o,\,k_o)] && < 0 \qquad \mbox{for focus}.
\end{eqnarray}

\end{document}